\documentclass[sn-mathphys-num]{sn-jnl}

\usepackage{graphicx}%
\usepackage{multirow}%
\usepackage{amsmath,amssymb,amsfonts}%
\usepackage{amsthm}%
\usepackage{mathrsfs}%
\usepackage[title]{appendix}%
\usepackage{xcolor}%
\usepackage{textcomp}%
\usepackage{manyfoot}%
\usepackage{booktabs}%
\usepackage{algorithm}%
\usepackage{algorithmicx}%
\usepackage{algpseudocode}%
\usepackage{listings}%

\begin{document}

\title[Article Title]{Generating Millions Of Lean Theorems With Proofs By Exploring State Transition Graphs}

\author*[1]{\fnm{David S.} \sur{Yin}}\email{dyin126@student.fuhsd.org}

\author[2]{\fnm{Jing} \sur{Gao}}\email{jinggao@purdue.edu}

\affil*[1]{\orgname{Lynbrook High School}, \orgaddress{\street{1280 Johnson Ave}, \city{San Jose}, \postcode{95129}, \state{CA}, \country{USA}}}

\affil[2]{\orgname{Purdue University}, \orgaddress{\street{465 Northwestern Ave}, \city{West Lafayette}, \postcode{47907}, \state{IN}, \country{USA}}}

\abstract{Large Language Models (LLMs) have demonstrated significant potential in generating mathematical proofs. However, a persistent challenge is that LLMs occasionally make mistakes, while even a minor mistake can invalidate an entire proof. Proof assistants like Lean offer a great remedy. They are designed for verifying each step of a proof in a formal language, and in recent years researchers have created AI models  to generate proofs in their languages. However, the scarcity of large-scale datasets of Lean proofs restrict the performance of such Automated Theorem Proving (ATP) models.

We developed LeanNavigator, a novel method for generating a large-scale dataset of Lean theorems and proofs by finding new ways to prove existing Lean theorems. By leveraging an interactive Lean client and an efficient method for proof step generation, LeanNavigator efficiently produces new theorems with corresponding proofs. Applying this approach to Mathlib4, we generated 4.7 million theorems totaling 1 billion tokens, surpassing previous datasets by more than an order of magnitude. 
Using this extensive dataset, we trained an AI model that outperforms the state-of-the-art ReProver model in theorem-proving tasks. These results confirm our hypothesis and demonstrate the critical role of large datasets in improving the performance of automated theorem provers.
}


\maketitle

\section{Introduction}\label{sec1}

In recent years we have witnessed the incredible growth in the capabilities of LLMs in a large variety of tasks, ranging from writing songs to creating spreadsheets. There have been many studies on using LLMs to create mathematical proofs \cite{schulz_2002, kovacs_voronkov_2013, polu_2020, lample_2022, yang_swope_2023}. But one major challenge is that an LLM occasionally makes mistakes in the proofs, while even a small mistake can invalidate the whole proof. This severely impedes LLMs from providing complex math proofs. For example, GPT-4o only solved 12\% of problems on the 2024 American Invitational Mathematics Examination (for high school students) \cite{openai_2024}.

This illustrates a key challenge in using generative models in solving mathematical problems, --- the inability to verify each step of its proof. Proof assistants such as Lean \cite{moura_ullrich_2021}, Coq \cite{huet_kahn_paulin_1997}, and HOL \cite{gordon_melham_1993} can naturally help, because they are designed exactly for this purpose. If a mathematical proof is written in the language of a proof assistant, the assistant can verify the correctness of each step of the proof. 

Among various proof assistants, Lean has emerged as a popular choice due to its intuitive design. It allows users to introduce hypotheses or subcases and offload much of the work to automated proof steps called tactics, simplifying the process of theorem proving. Lean has received great attention in recent years, with one prominent example being Terrence Tao, Timothy Gowers, Ben Green, and Freddie Manners proving the Polynomial Freiman-Ruzsa (PFR) conjecture using Lean in late 2023 \cite{gowers_green_tao_2023}. In 2024 DeepMind introduced AlphaProof, a Lean-based system employing reinforcement learning, which nearly achieved a gold medal in the International Mathematical Olympiad (IMO) 2024, alongside AlphaGeometry \cite{deepmind_2024}.

\begin{figure}[h]
	\centering
	\includegraphics[width=0.5\textwidth]{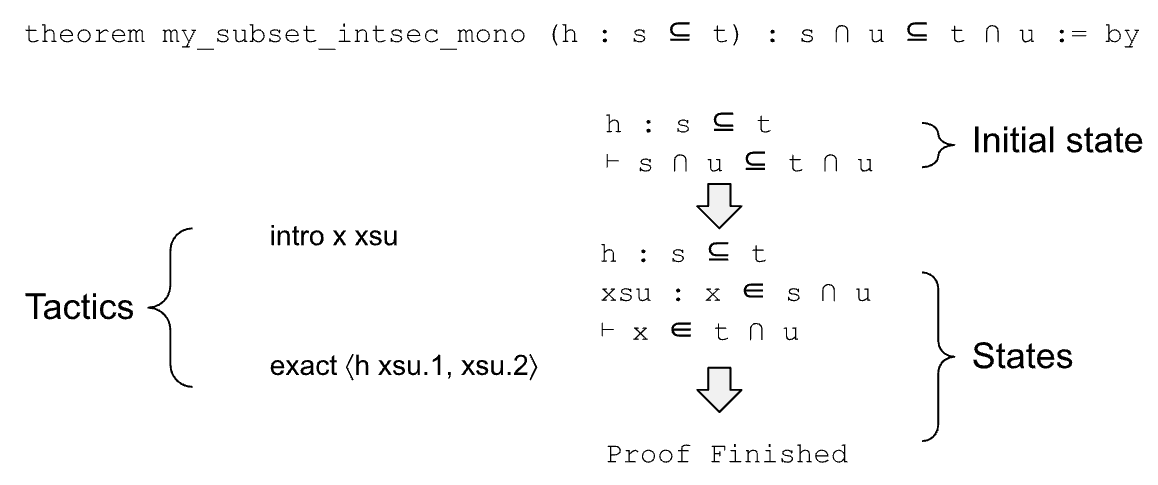}
	\caption{Lean proof of a very simple theorem ``\texttt{(h: s $\subseteq$ t) : s $\cap$ u $\subseteq$ t $\cap$ u}''. There is a hypothesis \texttt{h} which says ``\texttt{s $\subseteq$ t}'', and the target is to prove ``\texttt{s $\cap$ u $\subseteq$ t $\cap$ u}''. The proof contains two tactics. The first tactic ``\texttt{intro x xsu}'' introduces a new variable \texttt{x} and new hypothesis ``\texttt{xsu : x $\in$ s $\cap$ u}'', i.e., \texttt{x} belongs to the left-hand side of the target. It transforms the initial state into ``\texttt{h: s $\subseteq$ t, xsu : x $\in$ s $\cap$ u $\vdash$ x $\in$ t $\cap$ u}''. 	The second tactic ``\texttt{exact $\langle$ h xsu.1, xsu.2 $\rangle$}'' first applies \texttt{h} and \texttt{xsu.1} (i.e., ``\texttt{s $\subseteq$ t}'' and ``\texttt{x $\in$ s}'') on the target, and then applies \texttt{xsu.2} (i.e., ``\texttt{x $\in$ u}'') on the target, which finishes the proof.}
	\label{fig:lean-proof}
\end{figure}

However, proof assistants are only designed to verify proofs, not for generating proofs. A proof assistant is analogous to the Python runtime, which can execute a sequence of Python commands and return the state after each command. There is a need to create an ML model that can automatically generate a proof for a theorem, just like Copilot generating python code for a programming task. Figure \ref{fig:lean-proof} shows an example theorem and its proof.

In recent years the explosive growth of LLMs has encouraged researchers to explore their potential in training Automated Theorem Prover (ATP) models \cite{bansal_loos_2019, paliwal_loos_2020, polu_2020, yang_swope_2023}, which aim at generating proofs for Lean theorems. A major challenge, however, is the limited availability of large datasets for training these models. For example, ReProver \cite{yang_swope_2023}, a Lean-based automated prover published very recently, was trained on 112K theorems from Mathlib4. Although Mathlib4 is the official library of mathematical proofs of Lean and includes almost all major areas of mathematics, it still falls short of the scale required for optimal LLM training. 

In contrast, LLMs like Llama-2 have been trained on datasets containing trillions of tokens. It is impractical to expect an LLM to have good performances after being trained on such a small dataset. On the other hand, the existing corpus of math problems and solutions are not useful for training Lean proof models, because the proof for Lean (or other proof assistants such as Coq and Agda) is very different from a typical math proof. For example, a proof for a simple theorem such as ``\texttt{a * 1 = a}'' requires four special tactics in Lean (more details in Section 2.3). In fact LLMs such as GPT-4o understand Lean and can explain Lean theorems very well. But they have disappointing performances on generating proofs in Lean \cite{polu_2020, yang_swope_2023}.

Therefore, it is highly desirable to have a large dataset of Lean proofs, in order to train an LLM to generate math proofs in Lean automatically. In this paper we propose an approach that can generate millions of theorems in Lean, by exploring the state graphs of Lean theorems. As shown in Figure \ref{fig:lean-proof}, we can use a tactic to transfer a state into another one. We define a State Graph to be a directed graph of states, with a tactic on each edge (as shown in Figure 2). If we can identify many relevant tactics for each state, we can generate many new states from a given theorem. If there is a path from a new state to a ``Proof Finished'' node in the state graph, this new state can be considered as a new theorem, with a proof corresponding to the sequence of tactics on the path.

We introduce LeanNavigator, a method for generating a large number of theorems in Lean, by leveraging an interactive Lean client to traverse the state graphs of theorems. In order to efficiently generate relevant tactics and traverse the state graph, we use an embedding-based-retrieval method for tactic generation, which is more efficient than the generative model used by \cite{yang_swope_2023}. By applying LeanNavigator on Mathlib4, we generated 4.7M theorems in Lean, with 1B tokens in total. In contrast, the dataset used to train ReProver \cite{yang_swope_2023}, the previous state-of-the-art automated prover in Lean, contains only 57M tokens.

The larger dataset enables us to train a better model for automated proof generation, which outperforms ReProver \cite{yang_swope_2023} in two testing datasets.

Here are the list of main contributions made in this paper:

\begin{itemize}
\item We proposed LeanNavigator, a system that explores the state graphs of existing lean theorems, in order to generate many new theorems with proofs.
\item We generated 4.7M theorems with proofs by applying LeanNavigator on Mathlib, the official math library of Lean.
\item We trained a large language model using our dataset, which outperforms existing state-of-the-art models in proving theorems. This shows how a large dataset can help in training LLMs in generating mathematical proofs for Automated Theorem Proving.
\end{itemize}

The remainder of this paper is organized as follows. First, we will discuss the history of automated theorem proving and related datasets. Then, we will describe the format of our dataset and how to generate it. Finally, we will share our experimental results.

\section{Related Work}

\subsection{Proof assistants}\label{ssec:proofassist}

Over time, many softwares have been created to algorithmically check the correctness of mathematical proofs written in formalized mathematics, or a machine-readable language. Most proof assistants are based on type theory, which groups mathematical objects into types like integers and functions. 

In most proof assistants, proving a theorem means transforming a beginning state to an end state using tactics, or preexisting theorems. Each state could also be considered a theorem of its own. For example, the state ``\texttt{(a b c : $\mathbb{R}$) (a * b) * c = a * (b * c)}'' is the associative property of multiplication.

An early example of a proof assistant is HOL, which allows theorems to be proven in higher-order logic. However, it is fairly difficult to learn. Currently, the most popular proof assistants are Lean \cite{moura_ullrich_2021}, Agda \cite{bove_dybjer_norell_2009}, Coq \cite{huet_kahn_paulin_1997}, and Isabelle \cite{paulson_isabelle_1994}. Among them, Lean has gained the most popularity in recent years. Lean allows the user to introduce hypotheses or subcases, and a large amount of work can be offloaded to these tactics, making automated theorem proving much easier. In late 2023 Terrence Tao, Timothy Gowers, Ben Green, and Freddie Manners proved the polynomial Freiman-Ruzsa (PFR) conjecture using Lean \cite{gowers_green_tao_2023}. In 2024 DeepMind unveiled the Lean-based AlphaProof \cite{deepmind_2024}, which was close to getting a gold medal in IMO 2024, together with AlphaGeometry.   

\subsection{Automated Theorem Provers}\label{ssec:atp}
Currently, there are many models that attempt to perform automated theorem proving using methods like premise selection, which selects several potential tactics, and chooses one to apply. Recently, language models like GPT-f \cite{polu_2020} leveraging the power of generative models have also become viable for ATP. 

The HOList \cite{bansal_loos_2019} environment represents a significant step in applying machine learning to automated theorem proving in the primitive HOL formal logic language, as it was one of the earliest neural network based approaches. HOList integrates neural networks and reinforcement learning to guide the proof search process. Bansal et al. \cite{bansal_loos_2019} have also created a comprehensive dataset derived from formal proofs in the HOL Light system, capturing proof states, tactics, and other relevant information. 

Paliwal et al. (2019) \cite{bove_dybjer_norell_2009} investigates the use of graph neural networks (GNNs) to improve HOL theorem proving. The authors propose several graphical representations of higher order logic, including abstract syntax trees (ASTs), which are a tree representation of formal logic expressions. Each logical expression, including function applications, variable bindings, and operators, is encoded as a subtree within a larger AST. Nodes in an AST are variables, variable operators like functions, function definitions, and constant values. The study demonstrates that GNNs can effectively guide the proof search process by learning from these graph representations, leading to significant performance improvements in theorem proving tasks.

A very recent study employing large language models and Lean is ReProver \cite{yang_swope_2023}, which performs premise generation by first embedding the theorem state, and then choosing the premises with embeddings closest to the theorem's. ReProver's training data is obtained via LeanDojo \cite{yang_swope_2023}, a program that enables interaction with and extracts specific theorem data from the Lean formal logic environment. Other methods like PACT \cite{han_pact_2021} put more emphasis on the training format, trying methods like predicting masked parts of a proof. 

\subsection{LLMs As Automated Theorem Provers}\label{ssec:llmatp}
In recent years we have witnessed the improvements of LLMs in solving generic math problems. However, the generic purpose LLMs have disappointing performances on generating proofs in formal languages such as Lean. This is because generating mathematical proofs in a formal language is a distinct challenge, differing significantly from proving theorems in natural language. Consider the example from Lean's official tutorial (Mathematics in Lean), which proves ``\texttt{a * 1 = a}'', where \texttt{a} belongs to a group \texttt{G}.

In a typical mathematical book, one can probably prove the above theorem directly using the group's property. For example, here is a proof given by GPT-4o:

\vspace{0.1in}

\begin{quote}

``To prove the statement $a \cdot 1 = a$ where a belongs to a group, we rely on the group axiom of Identity element: There exists an element $1 \in G$ such that for every $a \in G$, $a \cdot 1 = 1 \cdot a = a$.

\vspace{0.1in}

\noindent By this axiom of the group, the identity element 1 satisfies $a \cdot 1 = a$ for all $a \in G$. Therefore, this equality holds directly from the definition of the identity element in a group.''
\end{quote}

\vspace{0.1in}

In contrast, below is the proof from \emph{Mathematics In Lean}, the official tutorial of Lean's Mathlib.

\vspace{0.1in}

\begin{quote}
    \hspace{-0.3in}
        \fbox{\begin{minipage}{3.25in}
        \small
            \texttt{theorem mul\_one (a : G) : a * 1 = a := by} \\
            \texttt{rw [$\leftarrow$ mul\_left\_inv a]} \\
            \texttt{rw [$\leftarrow$ mul\_assoc]} \\
            \texttt{rw [mul\_inv\_cancel]} \\
            \texttt{rw [one\_mul]}
        \end{minipage}}
\end{quote}

\vspace{0.1in}

We can see the proof in Lean is much more complex. Mathlib contains a theorem saying ``\texttt{1 * a = a}'', we must convert the above theorem into that. However, four steps are needed because group multiplication is noncommutative. The first step rewrites the ``\texttt{1}'' into ``\texttt{(a$^{-1}$ * a)}'' using the definition of a multiplicative inverse. The second step removes the parentheses using the associative property of multiplication. The third step uses the multiplicative inverse definition again, turning the state into ``\texttt{1 * a = a}''. The final step proves the theorem using the fact that multiplication between the identity element of a group and any other element of it keeps it the same. 

From the above example we can see that, although LLMs have been trained on huge datasets and could finish some relatively easy proofs in natural languages, such training is not very helpful for generating proofs in Lean.

\section{Methodology}

\subsection{State Graph of Lean}

In this section we will describe the format of our dataset and the generation process. Before describing our method, we would first show an example of a lean theorem and its proof. Below is a very simple theorem stating that a * b * c = b * (a * c), where a, b and c are real numbers. 

\vspace{0.1in}

\begin{quote}
	\fbox{\begin{minipage}{6in}
			\texttt{theorem my\_mul\_comm\_assoc (a b c : $\mathbb{R}$) a * b * c = b * (a * c) := by} \\
			\texttt{rw [mul\_comm a b]} 	   // Note: current state=``\texttt{b * a * c = b * (a * c)}'' \\
			\texttt{rw [mul\_assoc]}       // Note: current state=``\texttt{b * (a * c) = b * (a * c)}'', ProofFinished
	\end{minipage}}
\end{quote}

\vspace{0.1in}

At the beginning of the proof, we have the initial state of the theorem, which is:

\begin{quote}
	\fbox{\begin{minipage}{2.2in}
		\texttt{a b c : $\mathbb{R}$} \\
		\texttt{$\vdash$ a * b * c = b * (a * c)}
	\end{minipage}}
\end{quote}

At any step of a Lean proof, there is always a current state. The initial state is simply the theorem itself: ``\texttt{a * b * c = b * (a * c)}''. A proof consists of a list of tactics, each transits the current state into another state, which becomes the new current state. For example, the first tactic ``\texttt{rw [mul\_comm a b]}'' applies the commutative property of multiplication on ``\texttt{a}'' and ``\texttt{b}'', which transfers the initial state into a new state ``\texttt{b * a * c = b * (a * c)}''. The second tactic ``\texttt{rw [mul\_assoc]}'' applies the associative property, which transfers the state into ``\texttt{b * (a * c) = b * (a * c)}'', which is tautology and finishes the proof.

We can draw an analogy between a Lean proof and a Python program. A theorem is like a programming task, with variables defined in the Python environment. Each tactic is like a Python command, which modifies or creates some variables, leading to a new ``state'' in Python. The task is finished when the current state satisfies certain requirements (e.g., tautology).

\begin{figure}[h]
	\centering
	\includegraphics[width=1.0\textwidth]{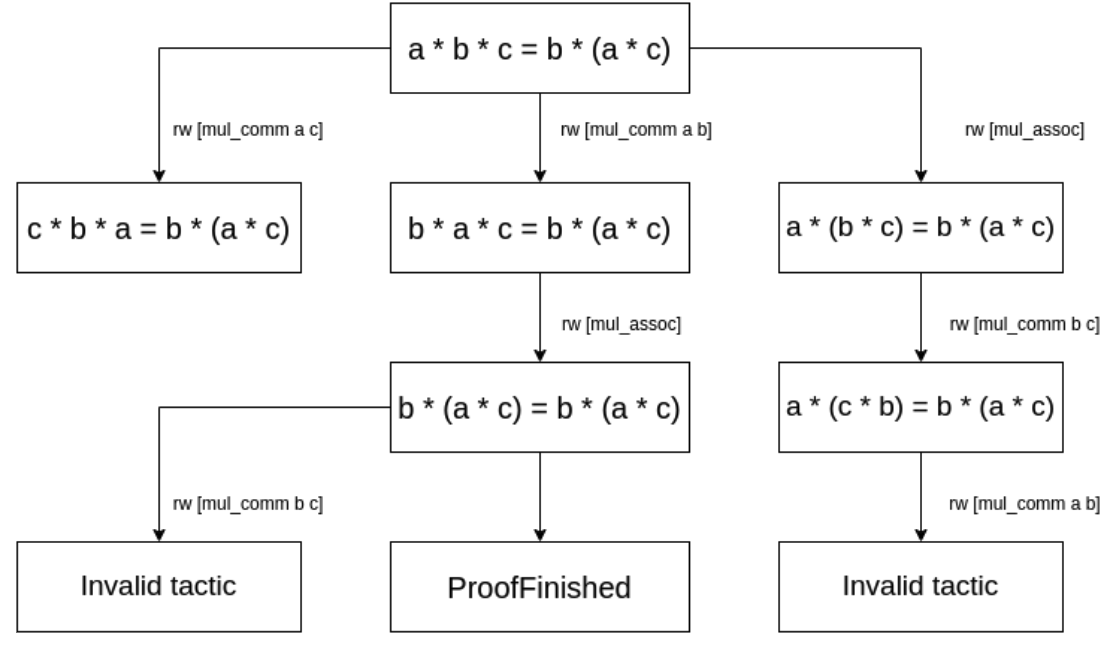}
	\caption{An example Lean state graph of theorem ``\texttt{theorem my\_mul\_comm\_assoc (a b c : $\mathbb{R}$) a * b * c = b * (a * c)}''. Please note a randomly chosen tactic is seldom applicable to a state, and therefore we need to train models to predict useful tactics.}
	\label{fig:state-graph}
\end{figure}

Figure \ref{fig:state-graph} shows how the initial state of theorem \texttt{my\_mul\_comm\_assoc} is transited with each tactic applied, and finally reaching the ProofFinished state. This can be considered as a directed graph of states, with a tactic on each edge.

Please note that the above theorem does not contain any hypotheses. In many theorems a state can contain one or more hypotheses. For example, the following state contains a hypothesis h, which says ``\texttt{a = b + c}''.

\begin{quote}
	\fbox{\begin{minipage}{3.1in}
			\texttt{a b c : $\mathbb{R}$} \\
			\texttt{h : a = b + c} \\
			\texttt{$\vdash$ a * a = b * b + 2 * b * c + c * c}
	\end{minipage}}
\end{quote}

\subsection{Dataset Description}

Each element in our dataset contains a theorem, which is created from a Lean state. It also contains a proof, which is a list of tactics that would transfer this theorem's initial state to ``\texttt{ProofFinished}''. This allows us to train a LLM to output a valid proof given a state. During the process of generating a proof, we can use Lean's interactive client to verify if each tactic generated is applicable to the current state. Therefore, to prove a theorem, we will use the LLM to repeatedly generate a new tactic, test each tactic using Lean's client, and stop when we reach a state of ``\texttt{ProofFinished}''. 

Below is an example of how a new theorem can be generated using tactics.

\begin{quote}
	\fbox{\begin{minipage}{6.2in}
			\texttt{theorem mul\_inv\_rev (a b : G) : (a * b)$^{-1}$ = b$^{-1}$ * a$^{-1}$ := by } \\
			\texttt{rw[$\leftarrow$ one\_mul (b$^{-1}$ * a$^{-1}$)]}             // Note: Current state: \texttt{(a * b)$^{-1}$ = 1 * (b$^{-1}$ * a$^{-1}$)} \\
			\texttt{rw[$\leftarrow$ mul\_left\_inv (a * b)]}            // Note: Current state: \texttt{(a * b)$^{-1}$ = (a * b)$^{-1}$ * (a * b) * (b$^{-1}$ * a$^{-1}$)}
	\end{minipage}
}
\end{quote}
Example 1: A more complex Lean theorem showing the utility of tactics
\vspace{0.1in}

\begin{quote}
	\fbox{\begin{minipage}{6in}
			\texttt{theorem mul\_inv\_rev\_proof (a b : G) : (a * b)$^{-1}$ = (a * b)$^{-1}$ * (a * b) * (b$^{-1}$ * a$^{-1}$) := by } \\
                \texttt{rw[mul\_left\_inv (a * b)]} \\
			\texttt{rw[one\_mul (b$^{-1}$ * a$^{-1}$)]} \\
                \texttt{rw[$\leftarrow$ one\_mul (b$^{-1}$ * a$^{-1}$)]} \\
                \texttt{rw[mul\_assoc]} \\
                \texttt{rw[mul\_assoc]} \\
                \texttt{rw[$\leftarrow$ mul\_assoc b b$^{-1}$]} \\
                \texttt{rw[mul\_right\_inv]} \\
                \texttt{rw[one\_mul]} \\
                \texttt{rw[mul\_right\_inv]} \\
                \texttt{rw[mul\_one]} \\
	\end{minipage}
}
\end{quote}
Example 2: The proof for the theorem generated by state navigation in Example 1
\vspace{0.1in}

In Example 1, the initial state says that for group elements \texttt{a} and \texttt{b}, the inverse of \texttt{(a * b)} is equal to the inverse of \texttt{b} multiplied by the inverse of \texttt{a}. is used to prove another more complex state. The first tactic uses the fact that for any group element \texttt{a}, the identity element multiplied by \texttt{a} does not change it. The second tactic uses the fact that a group element's inverse multiplied by the element results in the identity element. These two tactics can transform a state into a very different one, showing that traversing the state graph can lead to meaningful new states, which can be used as new theorems.

Example 2 provides a proof for the theorem derived in Example 1. It first applies the inverse tactics in Example 1, and then expands the right side of the equation using the associative theorem of multiplication and the fact that a group element multiplied by its inverse is the identity element. Finally, it simplifies the equation by canceling out group elements with their inverses.

\subsection{Tactic Generation}

Some readers may think that to prove a theorem in Lean, it would be feasible to try every single tactic in Mathlib. However, that is incredibly costly. Mathlib has more than 200 built-in tactics like \texttt{simp} and \texttt{rw}. Since most theorems in Mathlib like \texttt{one\_mul} and \texttt{mul\_assoc} can also be used as tactics, the total number of tactics can reach into the hundreds of thousands. As the number of possible states grows exponentially with the number of tactics, applying every tactic for every state is obviously infeasible. Therefore, a more efficient method is needed.

Let us first review how previous methods like ReProver \cite{yang_swope_2023} generate tactics. It contains two versions: Direct generation and RAG. The direct-generation version generates the next tactic based on the current state of the theorem, which is treated as a string. The RAG version first embeds the current theorem state by treating it as a string. Then it embeds a list of tactics whose embeddings are nearest to the theorem state's embedding to a list of potential tactics. These tactics are then fed with the theorem state into a language model to create a more accurate prediction.  

However, finding the right tactic can be very difficult, as Lean tactics can be passed different variables and hypotheses can be given arbitrary names. For example,  ``\texttt{rw [mul\_comm a b]}'' and 	``\texttt{rw [mul\_comm x y]}'' are considered two different tactics. In different types of proofs, the variable names are likely to be different even in two uses of the same tactic in the same proof, meaning that the tactics have different embeddings. This might not be a problem if there are billions of complex theorems to train on, but currently Lean's Mathlib library only contains around 100,000 theorems, with only 3.8 tactics each on average. 

To solve this problem, we convert tactics into tactic templates. Tactic templates have all variables replaced with placeholders. In the example in Figure 1, the first intermediate tactic would be changed to ``\texttt{rw [mul\_comm \{var0\} \{var1\}]}'', where \texttt{\{var0\}} and \texttt{\{var1\}} represent two different variables. We will also replace hypotheses like ``\texttt{a = b}'' or ``\texttt{k > 1}'' with \texttt{\{hypothesis\}} and unrecognized parts of our tactics with \texttt{\{unknown\}}. Then, we will embed the tactic template using a GPTNeo-350M model \cite{black_gpt_2021} trained to embed tactic templates and relevant theorem states to similar vectors. The model is trained using contrastive learning \cite{chen_kornblith_hinton_2020}, where each example contains a theorem state, a correct tactic, and a randomly selected incorrect one. The model then aims to embed the state as close to the correct tactic's embedding as possible.

This approach has two advantages. The first is that it will make identical embeddings for different usages of the same tactic, as variable and hypothesis names are standardized. The second advantage is that it requires less training data to understand the semantics of a tactic, as each tactic template appears more often than the raw tactics.

We place the embeddings of all tactic templates into a FAISS index \cite{douze_faiss_2024}, which enables nearest neighbor search based on embedding similarity. To generate tactics for a theorem state, we will first turn the state into an embedding using the previously mentioned model. We then find the 100 nearest tactic template embeddings from the FAISS index. Finally, we will generate all possible combinations of variables and hypotheses that the tactic could use. As the number of possible combinations grows exponentially with the number of templates, we will limit the number of generated tactics to 200.

\subsection{Dataset Generation}

One attractive property of Lean is that it allows efficient transitions among its states. In our experiments, the average time spent on applying a tactic was only 0.12 seconds per tactic. Given a theorem which has an initial state, we can keep applying tactics one by one, to transit among many different states. Each new state consists of a new set of hypotheses and/or a new goal, and this new state can be considered as a new theorem.

When generating our dataset, we will store information like the current state, the states leading to the current state, and the tactics required to prove the state from previous states. This allows us to construct a collection of graphs where nodes are states, edges are tactics, and the tactics in a path from one node to the other is a proof. We will call this graph the state transition graph. Therefore, to generate a proof, we have to traverse the state transition graph to find a path between the initial state and the proven state.

If a theorem is proven in a node, we will refer to it as a \texttt{ProofFinished} node. All \texttt{ProofFinished} nodes in the state graph of a theorem can be considered as a single node. If the \texttt{ProofFinished} node is reachable from a certain node $n$ in the graph, we call $n$ a provable node, and the corresponding state $s$ is a provable state. We can construct a theorem for each provable state, by collecting the tactics along the path from its node to the \texttt{ProofFinished} node, which is a proof for this theorem. Please note there can be multiple paths from a state to the \texttt{ProofFinished} node, each corresponding to a different proof. We pick the proof with the least number of tactics as our proof (and break ties by total string length of the tactics). 

In this study we built an efficient approach to explore this huge state graph, in order to generate millions of theorems, each of which being a distinct state. 

In the algorithm, we will run a breadth-first search from the initial state specified by the theorem to construct the state graph mentioned in Section 3.1.

\vspace{0.1in}
\noindent \fbox{\begin{minipage}{6in}
	Algorithm 1: ExploreStates(theorem)
	
	\hspace{1em} initial\_state $\leftarrow$ dojo.enter(theorem)
	
	\hspace{1em} state\_dict $\leftarrow$ \{initial\_state: [None, [], []]\}  \# state graph in our dataset
	
	\hspace{1em} push the initial\_state into the priority queue with priority 0 
	
	\hspace{1em} state\_idx $\leftarrow$ 1 
	
	\hspace{1em} while queue is not empty: 
	
	\hspace{2em} curr\_state $\leftarrow$ the state with lowest priority 
	
	\hspace{2em} pop the state with lowest priority 
	
	\hspace{2em} convert curr\_state into embedding with state-embedder 
	
	\hspace{2em} find tactic templates that are closest to above embedding from the faiss index 
	
	\hspace{2em} for each template: 
	
	\hspace{3em} substitute placeholders with variables and hypotheses to generate tactics
	
	\hspace{3em} for tactic in tactics: 
	
	\hspace{4em} try applying tactic to curr\_state to get next\_state 
	
	\hspace{4em} if next\_state is error: 
	
	\hspace{5em} continue 
	
	\hspace{4em} else: 
	
	\hspace{5em} state\_data$\leftarrow$(curr\_state,tactic,state\_dict[curr\_state].prefix + tactic)
	
	\hspace{5em} add state\_data to state\_dict[next\_state]
	
	\hspace{5em} if next\_state is not \texttt{ProofFinished}: 
	
	\hspace{6em} push next\_state into the priority queue with priority state\_idx 
	
	\hspace{6em} state\_idx += 1 
	\end{minipage}
}
\vspace{0.1in}

To run a tactic, we will use the LeanDojo \cite{yang_swope_2023} tool, which allows us to automatically apply tactics to a Lean state. 

Since the state graph of a theorem can be very huge or even infinite, we limit the amount of computation done with each theorem. We will terminate it after 30 minutes, or 200,000 state transitions. We will then add all states that have a distance of 8 tactics or less from a ProofFinished state into our dataset, each of which being a theorem with a proof consisting of at most 8 tactics. 

\section{Experimental Results}\label{sec2}

In this section, we will describe our evaluation process and share its results. All evaluation is performed on an NVIDIA A6000 with 48GB of memory, and an Intel i7-12700K. Evaluation is done using two benchmarks, the mathematics-in-lean(MIL) \cite{avigad_2020} and the MiniF2F benchmark \cite{zheng_minif2f_2021}. MIL is the official tutorial of Mathlib (Lean’s math library), consisting of many theorems covering advanced fields such as group theory, topology, and number theory. The following is an example theorem in the dataset, which involves profound knowledge of different tactics and mathematical experience. 

\vspace{0.1in}

\begin{quote}
    \fbox{\begin{minipage}{5.2in}
        \texttt{theorem natAbs\_norm\_mod\_lt (x y : gaussInt) (hy : $y \neq 0$) : (x $\mod$ y).norm.natAbs < y.norm.natAbs := by } \\
        \texttt{apply Int.ofNat\_lt.1} \\
        \texttt{simp only [Int.coe\_natAbs, abs\_of\_nonneg, norm\_nonneg]} \\
        \texttt{apply norm\_mod\_lt x hy} \\
    \end{minipage}}
\end{quote}

\vspace{0.1in}

The second dataset, MiniF2F, contains many competition math problems from sources like the AMC, AIME, and IMO. It includes problems like: 

\vspace{0.05in}

Which of the following is equivalent to 

$(2+3)*(2^2+3^2)*(2^4*3^4)*(2^8+3^8)*(2^{16}+3^{16})*(2^{32}+3^{32})(2^{64}+3^{64})?$

\vspace{0.05in}

Solving these problems requires creativity and intuition. For example, the above problem can be solved by multiplying $(3-2) = 1$, and applying the difference of squares.

\subsection{State Generation}

In order to generate a very large dataset, the speed of data generation is a key factor. Therefore, we first evaluate the efficiency of our method’s state generation, by running the dataset generation process in Section 3.3. For comparison, we will also use ReProver to generate states on the same dataset. We randomly select 121 theorems from the MIL dataset, let each approach navigate in the state graph, and count how many states each approach can reach within 2 minutes. The results are shown in Table 2, which shows the average number of states reached by each approach.  

\vspace{0.1in}

\begin{table}[b]
\begin{tabular}{|l|l|}
\hline
              & Average States / Theorem \\ \hline
LeanNavigator & 2035.45                  \\ \hline
ReProver      & 21.69                    \\ \hline
\end{tabular}
\vspace{0.1in}
\caption{A comparison between the state generation abilities of LeanNavigator and ReProver, with each method given two minutes per theorem.}
\label{table:state_gen}
\end{table}

As shown in Table \ref{table:state_gen}, LeanNavigator generates states far more efficiently than ReProver for the following two reasons:  

\vspace{0.1in}

(1) we use an embedding-based-retrieval based on FAISS(Douze et al. 2024) which is highly optimized, and 

\vspace{0.05in}

(2) ReProver uses a generative model which generates one byte at a time, while we retrieve the most relevant tactic-templates and fill the template with variable names.

\subsection{Dataset}

Having a large dataset can massively improve model performance, as shown by \cite{kaplan_scaling_2020}. For example, Llama \cite{touvron_llama_2023} was trained on 2T tokens of data. However, models like ReProver only have a small amount of data to work with (57M tokens). 

We use RAY with 24 processes, which work on up to 24 theorems simultaneously. It took 28 days to finish processing all the theorems in Mathlib4. Our dataset contains almost 20 times more tokens compared with ReProver’s dataset, as shown in Table \ref{table:dataset_size}, which enables us to train models with more prediction power without overfitting.

\begin{table}[]
\begin{tabular}{|l|l|l|}
\hline
              & Number of Theorems & Number of Tokens \\ \hline
LeanNavigator & 4.7M               & 1B               \\ \hline
ReProver      & 112K               & 57M              \\ \hline
\end{tabular}
\vspace{0.1in}
\caption{A comparison between LeanNavigator and ReProver's dataset sizes}
\label{table:dataset_size}
\end{table}

\subsection{Effectiveness In Training Predictive Models}

To measure the effectiveness of our dataset, we will also use it to train the flan-t5 base and flan-t5 small \cite{chung_scaling_2024} generative models, with 350M and 80M parameters respectively. We then compare our model with ReProver, \cite{yang_swope_2023}, which is the state-of-the-art method for generating lean proofs. 

ReProver uses the byt5-small model \cite{xue_byt5_2022}, a model with 300M parameters which treats each byte as a token in training and inference. This explodes the token length and is much slower than most LLMs using a tokenizer. It is probably OK for training on a small dataset, but is very expensive on a large dataset. Our model uses a BPE \cite{shibata_byte_pair_1999} tokenizer just like most language models today \cite{chung_scaling_2024}.  For example, the tactic \texttt{apply Int.ofNat\_lt.1} is tokenized into [\texttt{$<$s$>$}, \texttt{apply}, \texttt{Int}, \texttt{.}, \texttt{of}, \texttt{N}, \texttt{at}, \texttt{\_}, \texttt{lt}, \texttt{.}, \texttt{1}] by our tokenizer, while in ReProver each character is treated as a token.

We compare our model and ReProver \cite{yang_swope_2023} by using each of them to generate proofs on the MIL and MiniF2F benchmarks, in order to gauge their effectiveness across a wide range of fields. Please note that all our training data are created from the Mathlib4 dataset, and therefore our model has never seen any contents in MIL and MiniF2F. This is more rigorous than randomly sampling a part of the Mathlib4 dataset as the testing set (which was used in ReProver), because it contains many theorems that are similar to each other and have similar proofs. We only compare the performance of generative language models (instead of RAG models), since most popular LLMs today are generative models (such as Llama, Phi and Mistral).

Each model is given two minutes per theorem for evaluation, as otherwise the model may get into an infinite loop or continuously make meaningless transitions among the states. For each state, the model has ten tries to find a valid tactic that results in a new, unseen state. The evaluation will then continue from the first of those states. 

\vspace{0.1in}

\begin{table}[]
\begin{tabular}{|l|l|l|l|l|}
\hline
                & LeanNavigator flan-t5 base & LeanNavigator flan-t5 small & ReProver \\ \hline
MIL Results     & \textbf{39/117}        & 25/117              & 30/117   \\ \hline
MiniF2F Results & \textbf{104/493}       & 52/493              & 99/493   \\ \hline
\end{tabular}
\vspace{0.1in}
\caption{LeanNavigator and ReProver’s results on the MIL and MiniF2F benchmarks.}
\label{table:eval_results}
\end{table}

As you can see in Table \ref{table:eval_results}, the flan-t5 base model we trained on LeanNavigator significantly outperforms the ReProver model on MIL, and also outperforms it on MiniF2F. Our model has a similar number of parameters as byt5-small, which shows the effectiveness of our dataset and the importance of a large dataset when training LLMs for automated theorem proving.

\section{Conclusion}\label{sec13}

In this paper, we presented LeanNavigator, a novel approach to Lean dataset generation by exploring state transition graphs. Our method uses Lean’s tactic system to traverse the state graph, significantly expanding the dataset available for training automated theorem provers. We also introduce tactic templates in order to improve the efficiency in data generation. Through our approach, we generated 4.7 million theorems, — more than an order of magnitude larger than previous datasets like that used by ReProver \cite{yang_swope_2023}.

Our experiments demonstrate that LeanNavigator leads to notable improvements in automated theorem proving. The larger dataset allowed us to train a language model that outperforms existing state-of-the-art models in theorem proving benchmarks like the MIL \cite{avigad_2020} and MiniF2F \cite {zheng_minif2f_2021} datasets. 

These results underscore the importance of large-scale datasets in advancing the capabilities of automated theorem provers. LeanNavigator's ability to generate millions of diverse theorems provides the foundation for training new automated theorem proving models. We believe that our approach paves the way for even larger and more sophisticated models, enhancing the accuracy and reliability of proofs generated by automated systems. Future work could further optimize the tactic generation process and explore applications in other proof assistants, aiming to build a unified framework for formal theorem proving in mathematics.

\backmatter

\section*{Data availability}

The LeanNavigator dataset is hosted using Zenodo(a data repository by CERN) at \url{https://zenodo.org/records/13989482}. The code for dataset generation and testing are released at \url{https://github.com/davidsyin/leannavigator}. 

\section*{Declarations}
\subsection*{Conflict of Interest}
The authors declare that they have no competing interests.
\subsection*{Ethical Approval}
There are no ethical conflicts.
\subsection*{Funding}
The authors received no financial support for the research, authorship, and/or publication of this article.
\subsection*{Author's Contributions}
Both authors contributed to ideation and writing. David Yin contributed to implementation and experimentation.

\bibliography{lean-dataset-paper}
\end{document}